\documentclass[english,onecolumn,amsmath,amssymb]{revtex4-1}

\usepackage{comment}
\usepackage{graphicx}
\usepackage{color}
\usepackage{units}
\usepackage[normalem]{ulem}
\usepackage{docs}%
\usepackage{bm}%
\usepackage[colorlinks=true,linkcolor=blue,citecolor=red]{hyperref}%

\newcommand{\rv}{$\delta R^2 / R^2$ }
\newcommand{\res}{$R$ }

\begin{document}

\title{Effect of ambient on the resistance fluctuations of graphene}%

\author{Kazi Rafsanjani Amin}
\author{Aveek Bid}
\email{aveek.bid@physics.iisc.ernet.in}
\homepage[\\visit at:]{http://www.physics.iisc.ernet.in/~aveek_bid/aveek.html}
\affiliation{Indian Institute of Science, Bangalore, Bengaluru, Karnataka, India 560012}

\begin{abstract}
In this letter we present the results of systematic experimental investigations of the effect of different chemical environments on the low frequency resistance fluctuations of single layer graphene field effect transistors (SLG-FET). The shape of the power spectral density of noise was found to be determined by the energetics of the  adsorption-desorption of molecules from the graphene surface making it the dominant source of noise in these devices. We also demonstrate a method of quantitatively determining the adsorption energies of chemicals on graphene surface based on  noise measurements.  We find that the magnitude of noise is extremely sensitive to the nature and amount of the chemical species present. We propose that  a chemical sensor based on the measurement of low frequency resistance fluctuations of single layer graphene field effect transistor devices will have extremely high sensitivity, very high specificity, high fidelity and fast response times. 
\end{abstract}

\maketitle

The study of low frequency $1/f$ noise in graphene monolayer is interesting from both scientific as well as technological points of view. The specific surface area (2630 $m^2/g$) of single layer graphene (SLG) is amongst the highest in layered materials making the conductance of graphene extremely sensitive to the ambient - the presence of a few foreign molecules on its surface can significantly modify its electrical noise characteristics. The low defect levels of pristine graphene ~\cite{Novoselov22102004,doi:10.1021/nl080241l,doi:10.1021/es902659d,4773243,PhysRevLett.109.196601} ensures that intrinsic flicker noise due to thermal switching of defects are lower than any semiconductor material \cite{balandin2013low, xu2010effect, pal2011microscopic, kaverzin2012impurities,pellegrini20131}.
These  distinctly unique properties of single layer graphene make it exceptionally suited for use as chemical or radiation sensors.

Resistance fluctuation of pristine single layer graphene field effect transistor (SLG-FET) devices under high vacuum conditions have been studied in detail \cite{PhysRevLett.109.196601,pal2011microscopic, balandin2013low,4773243,xu2010effect}. There is considerable debate in the community as to which of the two possible mechanisms  is the dominant cause of noise in pristine SLG-FET devices~\cite{balandin2013low} - (1) mobility fluctuations due to charged scattering centers on substrate and device surface, or (2) number density fluctuations due to charged impurities on surface of device or on the substrate.  Although the effect of various chemical gas molecules on the resistance of the SLG-FET devices have been studied~\cite{amin2014graphene,yuan2013graphene, yavari2012graphene} there is very little study of the effect of exposure to different chemicals on the resistance fluctuations of SLG-FET devices. There has been a previous study of the effect of adsorbed molecules on resistance fluctuation spectrum~\cite{doi:10.1021/nl3001293} but a quantitative study of the energetics of the processes giving rise to the resistance fluctuations is missing. 

%%%%%% figure afmsem %%%%%%%%%%%%%%%%
\begin{center}
\begin {figure}[ht]
 \includegraphics[width=0.475\textwidth]{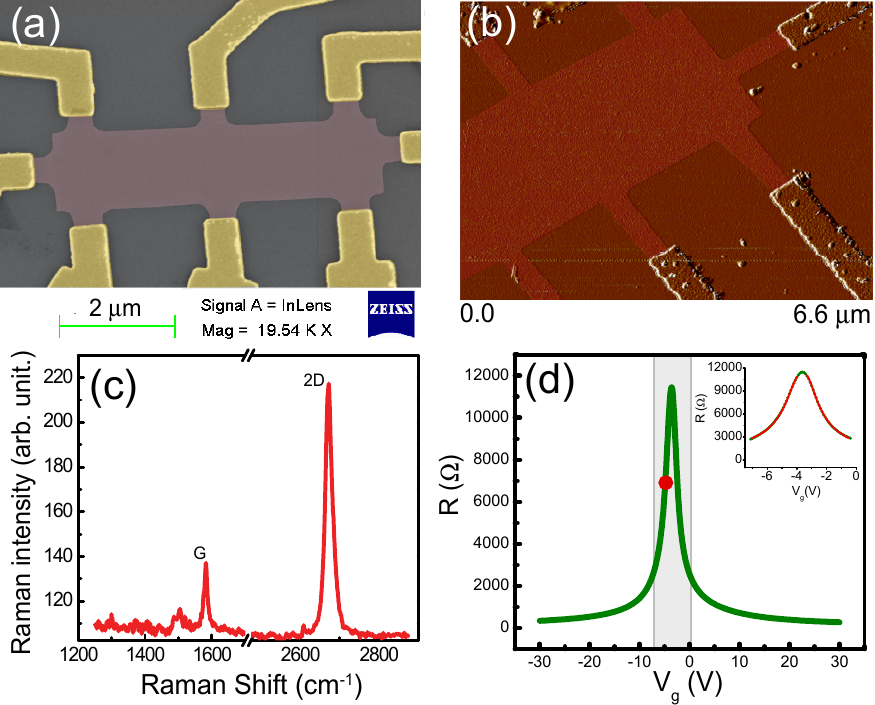} 
\small { \caption{ False color (a) SEM image and (b) AFM image of typical SLG-FET devices showing the cleanliness of the surface after the fabrication process. The surface roughness on the graphene extracted from AFM scan was 0.549 nm. (c) Raman spectrum of the SLG-FET device after lithography process. (d) Sheet resistance $R_{\Box}$ of the device as a function of back gate voltage $V_g$. The mobility of the device extracted from the data was 20,000~cm$^2V^{-1}s^{-1}$. Inset: Plot of $R-V_g$ data (shown in green circles) over a  narrow range of $V_g$. The red lines are the fit to the data using equation mentioned in reference ~\cite{PhysRevB.80.235402}. The fitting region is marked by gray box in the  main plot. \label{fig:images}}}
\end {figure}
\end{center}

SLG-FET devices were fabricated on $SiO_2$ substrates by mechanical exfoliation from natural graphite followed by conventional electron beam lithography process~\cite{Novoselov22102004}. 
 A false color scanning electron microscope (SEM) image of a typical device is shown in figure~\ref{fig:images}(a). The SLG is shown in violet while the yellow strips are the electrical contact lines made by thermally evaporating 5~nm of Cr and 70~nm of Au.

The devices were tested after the lithography process for the presence of resist residues using Atomic Force Microscope (AFM) and raman spectroscopy. A typical image of AFM scan and raman spectrum  is shown in figure~\ref{fig:images}(b) and figure~\ref{fig:images}(c) respectively. The surface roughness of the device estimated from the AFM line scans( $\sim$ 0.55~nm), and the  absence of a D peak and the position of the G peak (1582.2 cm$^{-1}$) in raman spectra indicates negligibly small extrinsic doping ~\cite{raman,PhysRevLett.97.187401,das2008monitoring,ferrari2007raman}. 
Quantum Hall measurements~\cite{Novoselov22102004} and Raman spectroscopy~\cite{PhysRevLett.97.187401} on representative devices were used to confirm that the graphene flakes were monolayers. The resistance of the devices were measured by standard low frequency ac techniques using a lock-in amplifier in a 4~probe configuration. A plot of the sheet resistance $R_\Box$ as a function of the back-gate voltage $V_g$ with the device in vacuum is shown in figure~\ref{fig:images}(d). The room temperature mobility of the devices were extracted from these measurements using the method described in reference~\cite{PhysRevB.80.235402}. 
The mobility values lay in the range 10,000- 26,000~cm$^2V^{-1}s^{-1}$ attesting to the high quality of the devices (See supplimental section)%\cite{kazibidsupple}.

\begin{center}
\begin {figure}[!t]
   \includegraphics[width=0.475\textwidth]{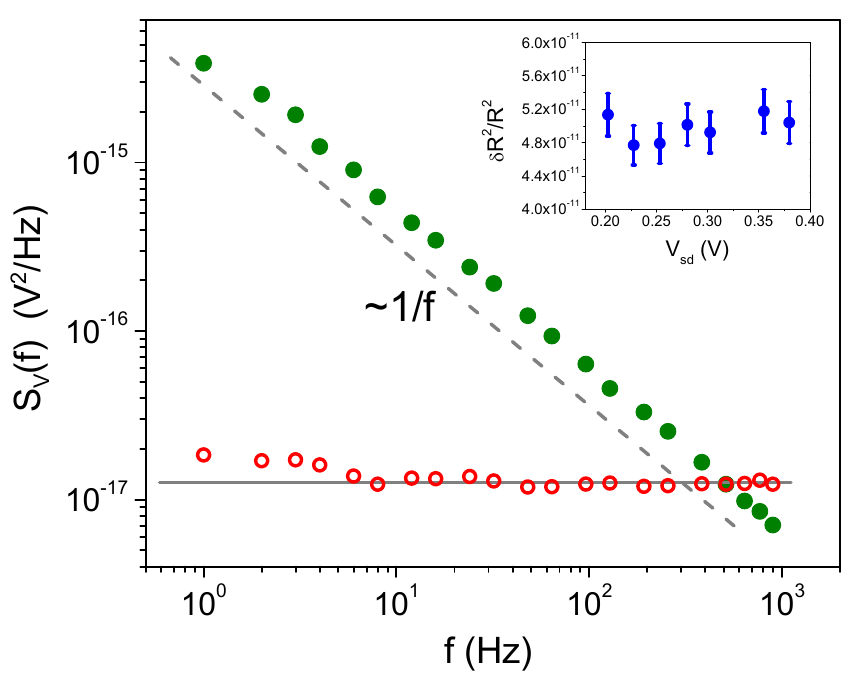}
      \small { \caption{A typical 1/f noise spectrum (olive filled circles) and background thermal noise (red open circles) of a pristine graphene monolayer FET device. The black solid line shows the expected background noise while the grey dotted line shows a reference 1/f curve. The inset shows the \rv as a function of $V_{sd}$. \label{fig:spectrum}}}
\end {figure}
\end{center}

The power spectral density (PSD) of voltage fluctuations $S_V(f)$ across the SLG-FET devices were measured as a function of frequency $f$ over a bandwidth of 1~Hz to 1~KHz  using an ac auto-correlation method [for details of the noise measurement and analysis process see ref~\cite{ghosh2004set,scofield1987ac}]. A typical PSD of pristine SLG-FET device measured at 295~K is shown in figure~\ref{fig:spectrum}. The red open circles are the measured background noise while the black solid line is the expected thermal noise for the device at 295~K. The excellent match between the two curves shows that the background noise arises primarily due to thermal noise of the device and that  extraneous instrumentation noise was negligible. The olive filled circles are the measured resistance fluctuation noise from the device (after subtracting out the background noise). It was seen that the PSD of pristine SLG-FET devices was always  $1/f$ in nature. For these (and all subsequent) measurements the chemical potential of the device was positioned  where the response of the sheet resistance $R_\Box$ to the  $V_g$ was  maximum (marked by the red dots in figure~\ref{fig:images}(d)).

%%% figure spectrum %%%%
%%%%%%%%%%%%%%%%%%%%%%%

%% figure separate, separate chloroform.... %%%%%
\begin{center}
\begin {figure}[ht]
    \includegraphics[width=0.475\textwidth]{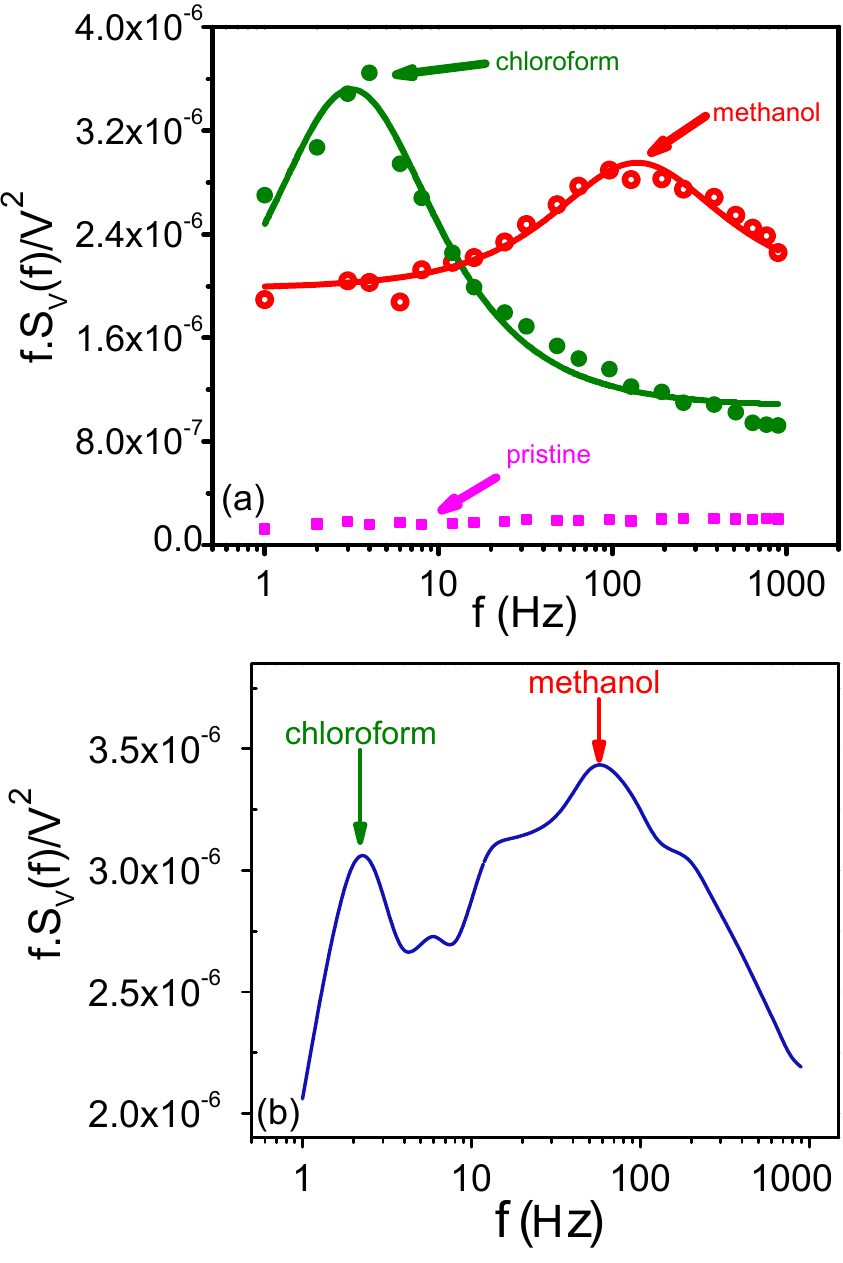}  
       \small { \caption{(a) Plot of scaled PSD as a function of frequency measured under different conditions - for the pristine device kept in vacuum (pink filled square), after exposure to 100 ppm methanol (red open circle) and after exposure to 100 ppm chloroform (olive filled circle). The solid lines are the fits to the experimental data using equation~\ref{eqn:svf}. (b) Plot of scaled PSD measured after the device was exposed to a mixture of 100~ppm methanol and 100~ppm chloroform - the characteristic frequencies of both methanol and chloroform  are marked by arrows. \label{fig:separate}}}
\end {figure}
\end{center}
%%%%%%%%%%%%%%%%%%%%%%%%%%%%%%%%%%%%%%%%%%%%%%%%%%

To quantify the effect of different chemical environments on the resistance fluctuations of SLG-FET devices we compare the PSD of the voltage fluctuations in the presence of different chemical species. An example is shown in figure~\ref{fig:separate}(a) where we plot the frequency dependence of the PSD of voltage fluctuations measured when the SLG-FET device is exposed to chloroform and methanol in two separate runs. For comparison we also show the normalized PSD of the pristine device. [Note that we have plotted the quantity $f\times S_V(f)/V^2$  - for $1/f$ noise the plot would be parallel to the frequency axis.] We find that the PSD shows Lorentzian humps at characteristic frequencies which allows us to fingerprint different chemicals. The data from these measurements were analyzed using an empirical relation that consisted of a $1/f$ term and a  Lorentzian with a characteristic component~\cite{PhysRevB.67.174415}

%%%% eqnarray svf
\begin{eqnarray}
S_V(f) = \frac{A}{f}+\frac{Bf_C}{f^2+(f_C)^2}
\label{eqn:svf}
\end{eqnarray}
%%%%%%%%%%%%%%%%

\noindent where $A$ and $B$ are constants 
are extracted from fits to the experimental data. The solid lines in figure~\ref{fig:separate}(a) are the fits to the experimental data using equation~\ref{eqn:svf}. 
The magnitude of the non $1/f$ noise component (quantified by the parameter $B$) contains information about the amounts of the chemical species the device has been exposed to while the shape of the PSD (parametrized by the Lorentzian of corner frequency $f_C$)  acts as unique spectroscopic fingerprints that helps identify the chemical.  An example of this is given in figure~\ref{fig:separate}(b) where we plot the the response of the device when it was exposed to a mixture of 100~ppm chloroform and 100~ppm methanol - the spectrum clearly shows the presence of both the chemical species. Such unique determination of analytes is obviously not possible in a detection scheme based on the measurement on resistance changes alone. The parameters $B/A$ and $f_C$ for 100 ppm of different chemical analytes extracted from our measurements is presented in table~\ref{tbl:fc} - the values of $f_C$ match very well with earlier reports~\cite{doi:10.1021/nl3001293}.

%%%%%%%%%%% table fc, b/a
\begin{table}[!h]
\caption{\label{tbl:fc}Values of the parameters  $f_C$ and $B/A$ for different chemical species}
\begin{ruledtabular}
\begin{tabular}{lll}
\textbf{Chemical species} & \textbf{$f_C$} & \textbf{B/A} \\
\itshape methanol & 138.68  & 0.977 \\ 
\itshape nitrobenzene & 82.0895  & 1.7582\\ 
\itshape chloroform & 3.156  & 4.571 \\ 
\itshape ammonia & 9.659  & 0.3942 \\ 
\end{tabular}
\end{ruledtabular}
\end{table}

The relative variance of the resistance fluctuations \rv (which we refer to as noise) was obtained by integrating the PSD $S_R(f)$ over the bandwidth of measurement:
%%%%%%%%%%%%%%%%%%%%%%%%%%%% equation relative variance
\begin{equation}
\frac{\delta R^2}{ R^2} = \frac{\int^{f_{max}}_{f_{min}}f  S_R(f) df}{R^2}
\end{equation}
%%%%%%%%%%%%%%%%%%%%%%%%%%%%%%%%%%%%%%%%%%%%%%%%%%%%%%%
 
\begin{center}
\begin {figure}[t]
  \includegraphics[width=0.475\textwidth]{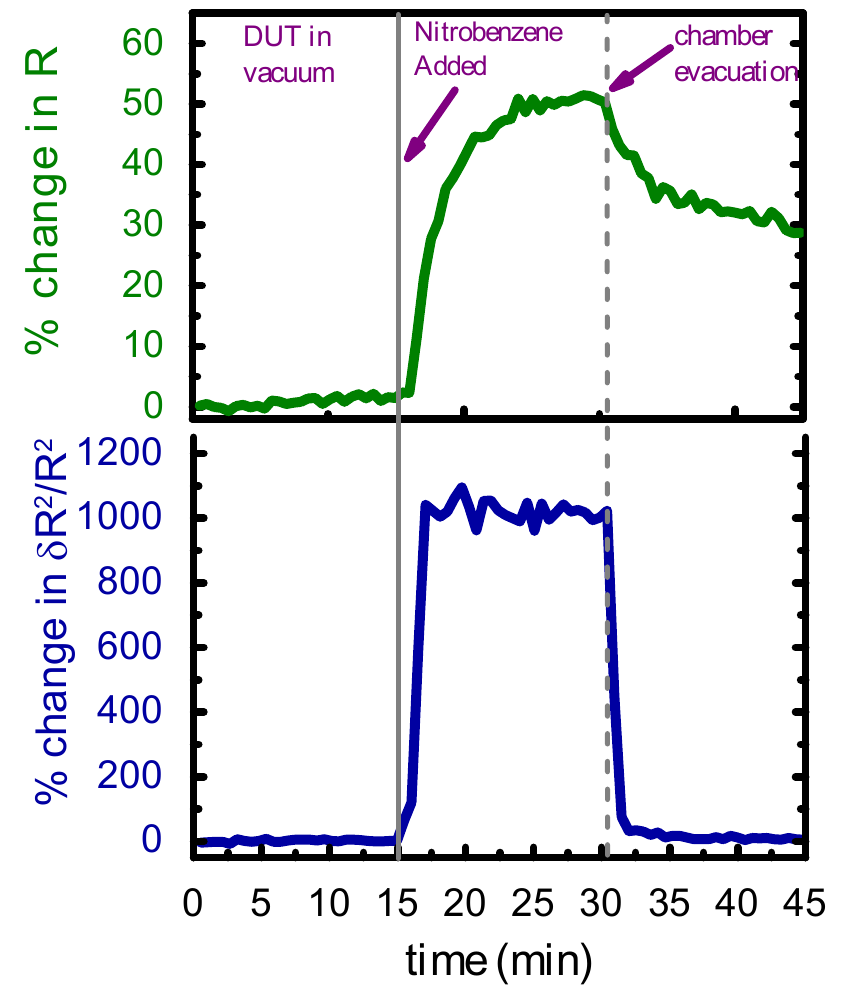}
      \small {\caption{Plot of change in \res (top panel, olive line) and change in \rv (bottom panel, blue line) with time for a typical measurement with the SLG-FET device exposed to 100~ppm of nitrobenzene.  \label{fig:compare}}}
\end {figure}
\end{center}

Figure ~\ref{fig:compare} shows the plot of the percentage changes in \res and \rv measured simultaneously  in a typical measurement where the device was exposed to  100~ppm of nitrobenzene vapor. For the initial 15 minutes the device was kept in vacuum and both the \res and \rv measured to establish the base values. As soon as nitrobenzene was introduced (at the instant of time marked by grey line) both  \res and \rv started increasing with time and eventually saturated. The change in \res was about 50~$\%$, whereas the change in \rv is 1000~$\%$, more than an order of magnitude higher.  The chamber evacuation was then started at the time shown in the figure as grey dotted line and both \res and \rv started decreasing  towards the baseline values. The resistance of the device took more than an hour to regain its initial value. This  large response time of the resistance change in graphene devices exposed to a chemical environment has been seen before~\cite{schedin2007detection, chen:053119,6475143} and is one of the major bottlenecks in implementing chemical sensors based on SLG-FET.  The noise of the device on the other hand resets to the baseline value within a few seconds of starting the chamber evacuation. We have performed similar measurements for different chemicals like acetone, methanol, chloroform and ammonia, and with eight different devices. The response of \res and \rv of the SLG-FET in all these cases were qualitatively similar  with the exact response depending on the type and amount of the chemical. From these measurements we can conclude that there are at least two major differences between the response of resistance and resistance fluctuations  of graphene devices to change in the chemical environment - (1) the magnitude of relative change in noise is much larger  and (2) the typical time scale associated with changes in noise is much smaller.

%%%%%%%%%%%%%%% figure comparison nitro
%%%%%%%%%%%%%%%%%%%%%%%%%%%%%%%%%%%%%%%

The drastically different trend of change in \res and in \rv upon exposure to a chemical environment can be understood from the following simple picture. The 
average resistance \res and resistance fluctuations \rv in  a sample arise from quite distinct mechanisms.
Static scatterers can have an appreciable effect on the resistance of a device while having negligible effect on its resistance fluctuations. On the other hand the presence of dynamic scatterers even with a  weak scattering potential can have a large affect on the resistance fluctuation spectrum while having very little effect on the average resistance~\cite{RevModPhys.60.537}.  Changes in the resistance of a graphene device
due to change in ambient conditions is directly related to the amount of the analyte molecules adsorbed  on its surface. The resistance of the device can go back to the baseline value only after  desorption of all the adsorbed molecules from the surface. This process is very slow for graphene, and might take up to few hours, depending upon the quantity and type of molecule adsorbed  \cite{schedin2007detection,chen:053119}.

Resistance fluctuations, on the other hand, arise primarily due to fluctuations in both the number density and mobility of the charge carriers in the system~\cite{balandin2013low}. For a semiconductor device exposed to a chemical environment there are three primary sources of resistance fluctuations: dynamic adsorption-desorption of the chemical species from the device surface, dynamic percolative motion of the chemical species on the device surface and charge trapping-detrapping by the chemical species~\cite{jokic2012fluctuations,meth3, Kish200055, meth1}.  Of these adsorption-desorption noise caused by fluctuations of the equilibrium number of adsorbed molecules in the device is predicted to dominate the resistance fluctuations seen in metallic sensors~\cite{meth3, Kish200055}. The adsorption-desorption process giving rise to the  resistance fluctuations in the device is facilitated by the presence of a reservoir of analyte  vapours close to the graphene surface. 
Pumping out the analyte vapours depletes this reservoir rapidly.
This slows down the adsorption-desorption process and this slow desorption continues till all the analytes have been removed from the device. A detailed analysis is presented in the supplementary material. % \cite{kazibidsupple}.

If adsorption-desorption noise is really the dominant source of resistance fluctuations in SLG-FET exposed to a chemical environment then $f_C$ should be related to the adsorption-desorption  energy $E_{a}$ of the specific gas molecule on SLG-FET through the equation~\cite{RevModPhys.60.537, PhysRevB.67.174415}:
%%%%%%%%%%%% equation fc lorenz
\begin{eqnarray}
f_C = f_0 exp \left(\frac{-E_a(T)}{k_BT}\right)
\label{eqn:fc}
\end{eqnarray} 
%%%%%%%%%%%%%%%%%%%%%%%%%%%%%%%%%%%%%%%
where $f_0$ is the attempt frequency for the thermally activated process.

%%%%%%%%%%%%%%%%% figure activation %%%%%%%%%%%%%%
\begin{center}
\begin {figure}[ht]
  \includegraphics[width=0.475\textwidth]{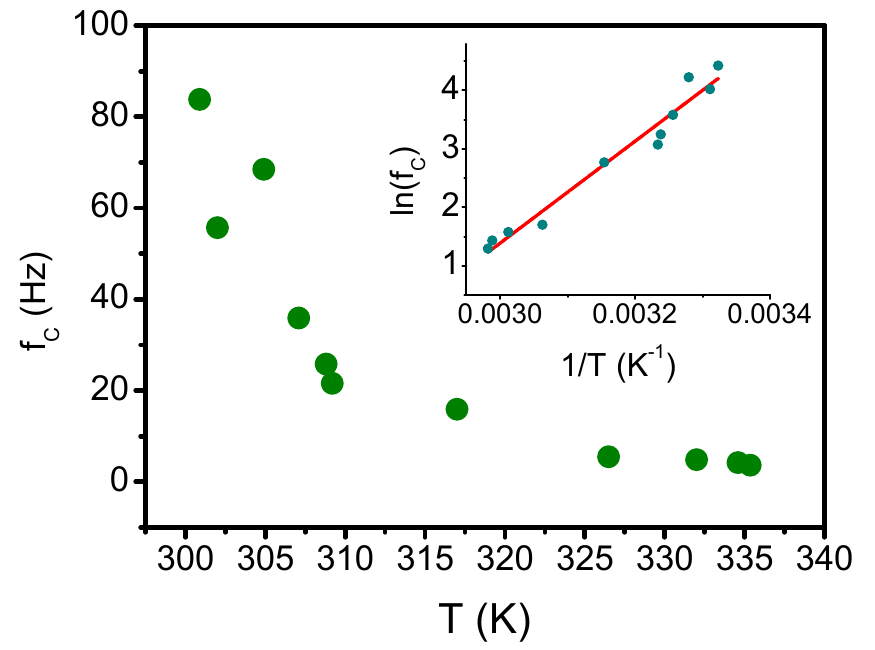}
    \small { \caption{Plot of $f_C$  as a function of temperature over the range of 300~K to 340~K. (inset) The filled circles show the plot of logarithm of $f_C$ as a function of inverse temperature, the red line is the fit to the data using equation~\ref{eqn:fc}.   \label{fig:activation}}}
\end {figure}
\end{center}
%%%%%%%%%%%%%%%%%%%%%%%%%%%%%%%%%%%%%%%%%%%%%%%
To test this hypothesis we have measured the temperature dependence of noise in SLG-FET devices in the presence of chemical vapors over the temperature range 300~K-340~K.  In figure~\ref{fig:activation} we show a plot of $f_C$ as a function of temperature extracted from these measurements - the data presented here have been obtained by exposing the SLG-FET device to 200~ppm of methanol. The inset shows a plot of the logarithm of $f_C$ as a function of inverse temperature, the red line is a least-square  linear fit to equation~\ref{eqn:fc}. The linearity of the data shows that the activation energy is temperature independent as expected over this narrow temperature range. The value of $E_a$ extracted from the slope of the curve is 752.3~meV~$\pm$ 6.30$\%$ which matches well the calculated values of adsorption energy of methanol on SLG-FET~\cite{methact1, methact2}.  
 
%\section{Conclusions}
To conclude, we have studied the effect of different chemical environment on the resistance fluctuations of single layer graphene FET devices. Our measurements indicate that the main source of noise in these devices is number density fluctuation arising from adsorption-desorption process of the chemicals at the graphene surface. We also find that a detection scheme based on the measurement of resistance fluctuations is far superior to the traditional method of measuring the average resistance change in terms of sensitivity, specificity and the response time of the detector.

%%%%%%%%%%%%%%%%%%%%%%%%%%%%%%%%%%%%%%%%%%%
%\begin{acknowledgments}
The work was supported by NPMASS, Aeronautical Development Agency (ADA), Govt. of India. The authors acknowledge device fabrication and characterization facilities in CeNSE, IISc, Bangalore. KRA thanks CSIR, MHRDG, India for financial support. KRA thanks Phanindra Sai for help and useful discussions on lithography.

%\end{acknowledgments}

%%%%%%%%%%%%%%%%%%%%%%%%%%%%%%%%%%%%%%%%%%%%%%%%%

%\bibliography{sensor}
\clearpage

\textbf{\huge{Supplimentary Section}}

%________________ transient behaviour
\section*{Mobility calculation}
Mobility and intrinsic doping concentration were extracted  by fitting the measured resistance \textit{vs}  gate voltage data to the equation ~\cite{PhysRevB.80.235402}
\begin{eqnarray}
R =  R_c +\frac{l}{w}\frac{1}{e\mu\sqrt{n_0^2+e^2C^2(V_g-V_d)^2}}
\label{eqn:fit}
\end{eqnarray} 
where $R_C$ is the contact resistance, $l$ and $w$ are channel length and width respectively, $\mu$ is the mobility, $n_0$ is the intrinsic doping concentration, $C$ is the gate capacitance per unit area, $V_g$ is gate voltage and $V_d$ is the Dirac point.  The fit to the equation has been included in figure~1(d) of main text - reproduced here for ready reference. The measurements for mobility extraction were performed with the device maintained in vacuum.
\begin{center}
\begin {figure}[!ht]
\includegraphics[width=.5\textwidth]{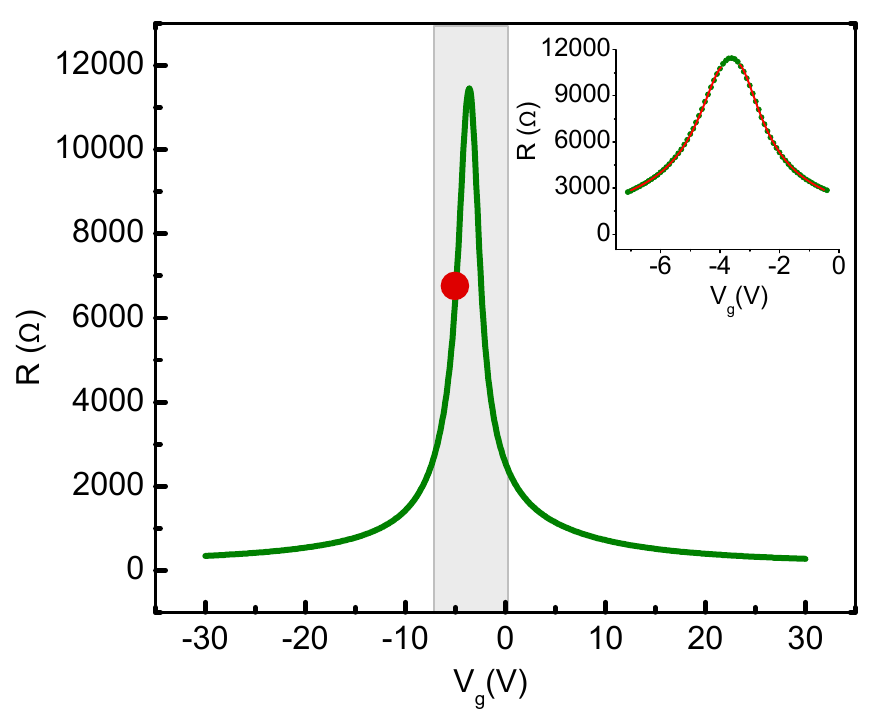}
\small { \caption{ False color (a) SEM image and (b) AFM image of typical SLG-FET devices showing the cleanliness of the surface after the fabrication process. The surface roughness on the graphene extracted from AFM scan was 0.549 nm. (c) Raman spectrum of the SLG-FET device after lithography process. (d) Sheet resistance $R_{\Box}$ of the device as a function of back gate voltage $V_g$. The mobility of the device extracted from the data was 20,000~cm$^2V^{-1}s^{-1}$. Inset: Same plot is shown, in green circles, in a  narrow range of $V_g$, and the red lines are the fit to the data using equation mentioned in reference ~\cite{PhysRevB.80.235402}. The fitting region is marked by gray box in the  main plot. \label{fig:rvg}}}
\end {figure}
\end{center}

\section*{Time scales associated with  changes in resistance and noise}

 The change in resistance (\res) and noise (\rv) of the device while adding analyte vapour or while pumping out the analyte vapour  shows transient behaviour with time.  While adding the analytes, the transient behaviour of the changes in \res and \rv with time can be described by the equation
\begin{eqnarray}
A(t) =  A_0 [1- e^{-t/\tau}]
\label{eqn:add}
\end{eqnarray}
 where we use $A$ to describe both  \res or \rv, in general.
The time constant $\tau$ is given by the negative of inverse of slope of the plot of $ln(1-A/A_0)$ with $t$, as given by equation~\ref{eqn:tau1}

\begin{eqnarray}
ln (1- A/A_0) = -t/\tau
\label{eqn:tau1}
\end{eqnarray}

\begin{center}
\begin {figure}[!ht]
\includegraphics[width=0.47\textwidth]{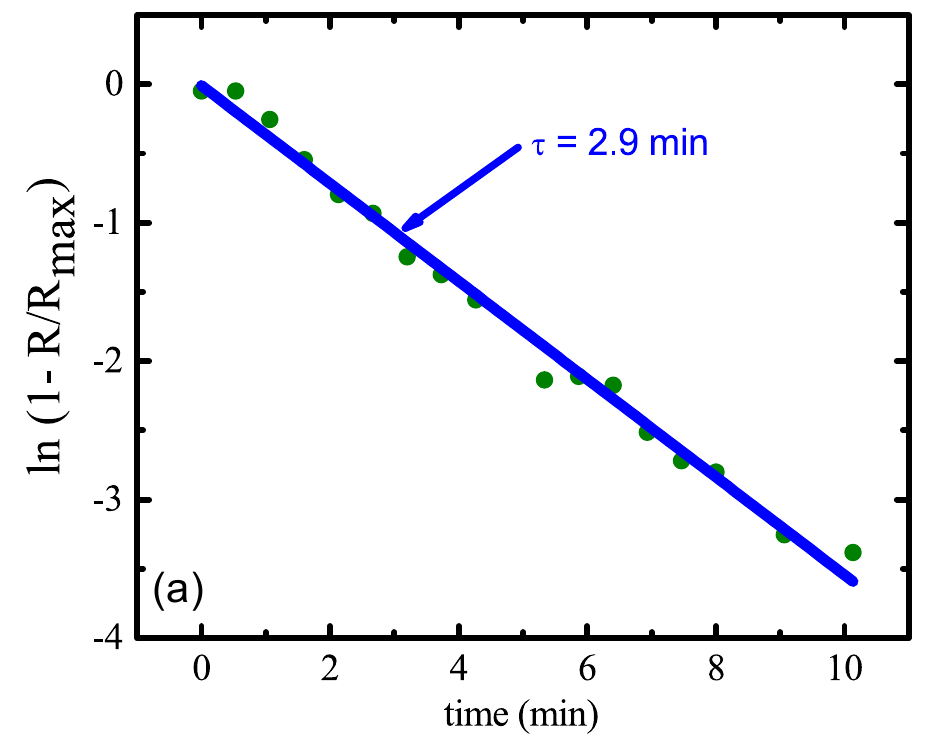}\hspace{2.5mm}\includegraphics[width=0.47\textwidth]{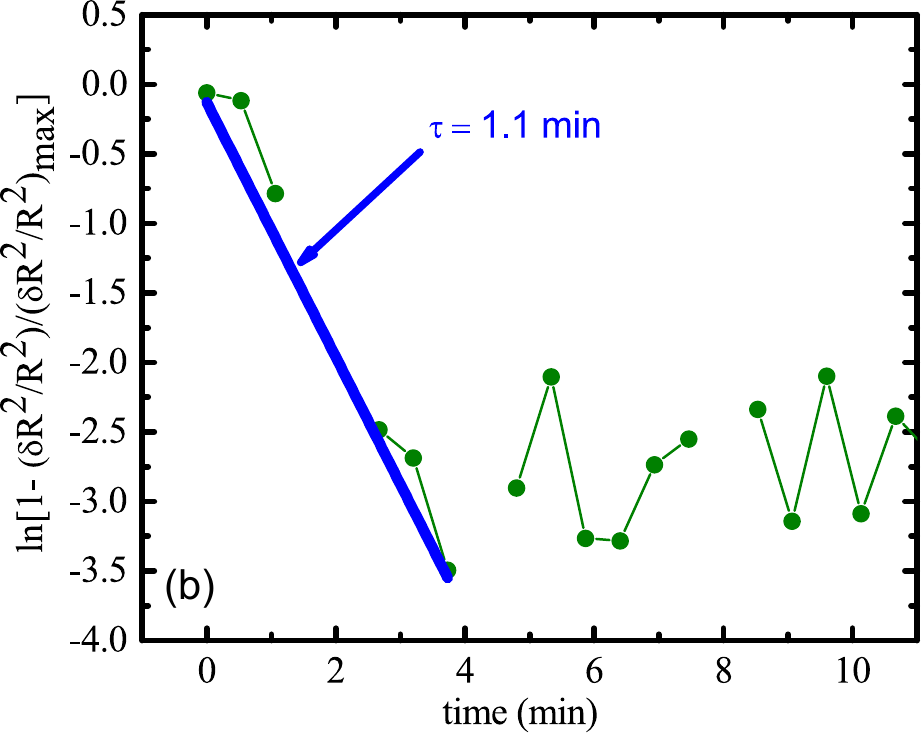}
\small { \caption{  Plot of (a) $ln(1-R_{max}/R_0)$ with $t$ and (b) ln[1- (\rv)/ (\rv)$_{max}$] with $t$ while nitrobenzene vapour was added, shown in green circles. The blue lines are the linear fit to the data. The starting of time axis is shifted to moment of addition of nitrobenzene.   \label{fig:add}}}
\end {figure}
\end{center}

Figure~\ref{fig:add}(a) shows a plot of $ln(1-R/R_{max})$ with $t$ while figure~\ref{fig:add}(b) shows a plot of ln[1-($\delta R^2 / R^2)/(\delta R^2 / R^2)_{max}$] with $t$. The solid lines are the linear fit to the data. The data is plotted for time starting from  the moment of adding of nitrobenzene. The values of the different time scales extracted from the fits ot the data are tabulated in table \ref{tbl:tauc}. The \rv of the device saturates much faster as compared to the resistance  as can be seen from the time constants extracted from the plots. 

%%_________________________ figure add ________________

Similarly the time variation of parameter $A$ while pumping the analytes out of the measurement chamber is described by the equation 
\begin{eqnarray}
A(t) =  A_0 e^{-t/\tau} 
\label{eqn:pump}
\end{eqnarray}
and the time scale associated with the desorption process was  calculated from 
\begin{eqnarray}
ln (A/A_0) = -t/\tau
\label{eqn:tau2}
\end{eqnarray}

%%________________ figure pump ________________
\begin{center}
\begin {figure}[t]
\includegraphics[width=0.475\textwidth]{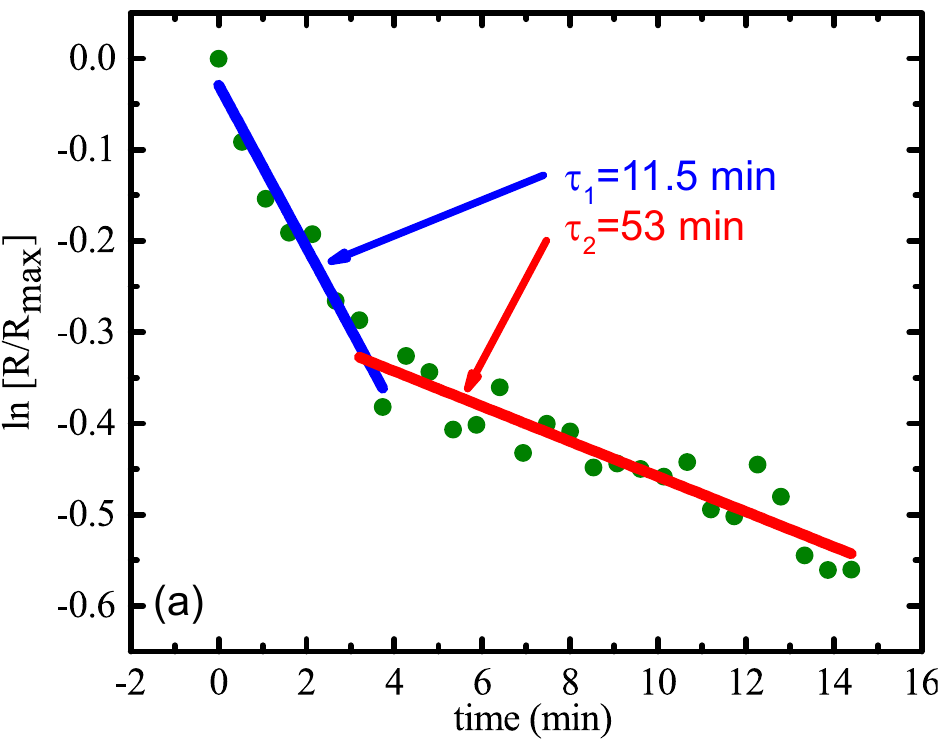}\hspace{2.5mm}\includegraphics[width=0.475\textwidth]{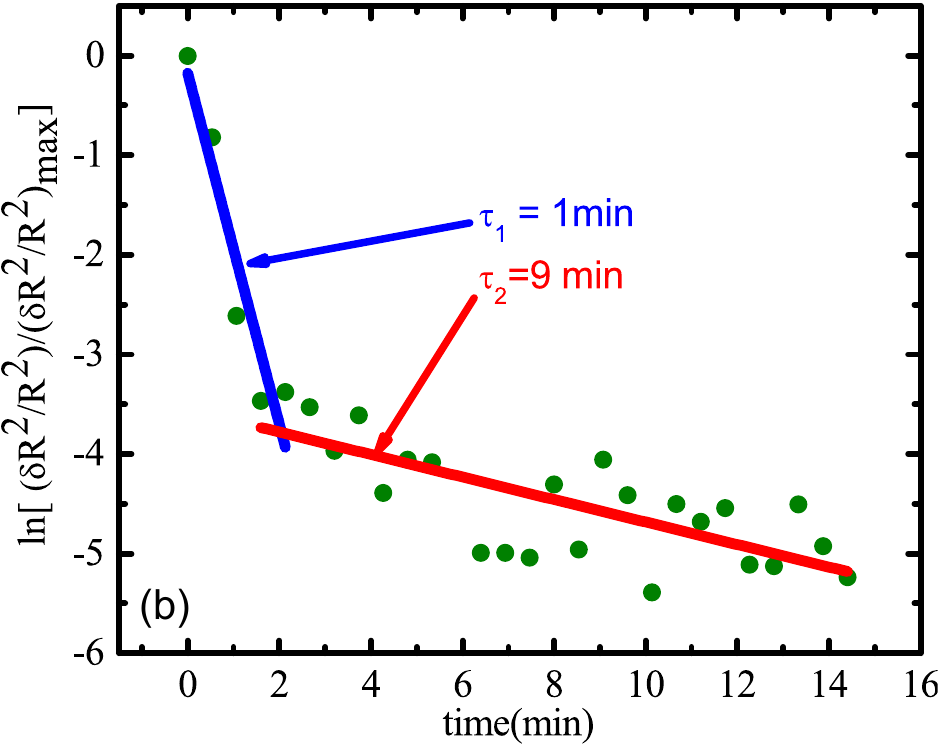}
\small { \caption{ Plot of (a) $ln(R_{max}/R_0)$  and (b) ln[(\rv)/ (\rv)$_{max}$] with $t$ while nitrobenzene vapour was added. The data is shown in green circles. The blue lines are the linear fit to the data, corresponding to the first time constant, and the red lines are the fit corresponding to the second slow time constant. The starting of the time axis has been shifted to the moment of the begining of the pumping. \label{fig:pump}}}
\end {figure}
\end{center}

Figure~\ref{fig:pump}(a) shows a plot of $ln(R/R_{max})$ with $t$ while figure~\ref{fig:pump}(b) shows a plot of ln[($\delta R^2 / R^2)/(\delta R^2 / R^2)_{max}$] with $t$. The solid lines are the linear fits to the data.
We have seen previously that during the addition of the analytes to the system, the rate of change of both the  resistance and the noise can be described by a single time constant (the values of the time constants for the two processes are different).  On the other hand during the pumping out process both resistance change and noise change have two different time constants - a faster tome constant in the initial 3-4 minutes of the process and a longer time constant later. Pumping out the analyte vapours depletes the reservoir of analytes rapidly. This slows down the adsorption-desorption process and this slow desorption continues till all the analytes have been removed from the device. As the number of such fluctuation in the time window of each measurement is now very small (compared to as it was before in presence of analytes in the reservoir) the frequency associated with fluctuation resulting from this process goes below the bandwidth of our measurement and the noise in the device reverts back very fast to the baseline value it had in the absence of analytes. The fast supression of dynamic absorprion-desorption due to depletion of the reservoir  followed by a slow desorption can be seen from the presence of two different time scales in the transient behaviours of \res and \rv.

Table~\ref{tbl:tauc} shows the various time constants extracted from logarithmic fits to the data shown in figure~\ref{fig:add} and figure~\ref{fig:pump}. Tentatively, we can attribute the first of these two time constants  to the depletion of the reservoir due to pumping out of the analytes and the second time constant to the slower desorption process of the analytes adsorbed on the graphene surface.  The first time scale corresponding to change in noise ($\sim$~1~min) is comparable to the time taken by the pumping system to achieve  the final vacuum in the measurement chamber.

%____________________________ table: time scale analysis ___________________
\begin{table}[!h]
\caption{\label{tbl:tauc}Values of the time constant during adding analyte}
\begin{ruledtabular}
\begin{tabular}{ll}
\textbf{Process} & \textbf{time constant (min)} \\
\itshape time constant for change in R during adding analyte & 2.9\\
\itshape time constant for change in $\delta R^2 / R^2$ during adding analyte & 1.1 \\ 
\itshape first time constant for change in R during pumping out of analyte & 11.5\\
\itshape second time constant for change in R during pumping out of analyte & 53 \\ 
\itshape first time constant for change in $\delta R^2 / R^2$ during pumping out of analyte & 1\\
\itshape second time constant for change in $\delta R^2 / R^2$ during pumping out of analyte & 9 \\ 
\end{tabular}
\end{ruledtabular}
\end{table}

\bibliography{sensor}

\end{document}